# Generation of picosecond pulses with variable temporal profiles and linear polarization by coherent pulse stacking in a birefringent crystal shaper

FANGMING LIU,[1] SENLIN HUANG,[1] SHANGYU SI,[1] GANG ZHAO,[1] KEXIN LIU,[1,*] AND SHUKUI ZHANG[2,*]

[1]*Institute of Heavy Ion Physics, School of Physics, Peking University, Beijing 100871, China*
[2]*Thomas Jefferson National Accelerator Facility, 12000 Jefferson Avenue, Newport News, VA 23606, USA*
[*]*Corresponding authors: kxliu@pku.edu.cn , shukui@jlab.org*

**Abstract:** We report the study and demonstration of a new laser pulse shaping system capable of generating linearly polarized picosecond laser pulses with variable temporal profiles including symmetric intensity distributions such as parabolic, flattop, elliptical, triangular, as well as non-symmetric distributions, which are highly desired by various applications. It is found that both high transmittance and high stability of the shaped pulse can be achieved simultaneously when crystals are set at a specific phase delay through fine control of the crystal temperature. Although multi-crystal pulse stacking with different configurations were reported before particularly for flattop pulse generation, this new configuration leads to new opportunities for many potential applications over a wide range of laser wavelengths, pulse repetition rate, time structures and power levels. A practical double-pass temporal shaping configuration that significantly reduces the number of crystals is also proposed in this paper as a result of this work.

## 1. Introduction

Temporal laser shaping is highly desirable in a wide variety of fields and applications. For example, laser pulses with a parabolic shape can be used to minimize the spectral ripple and wave breaking in super-continuum generation(SCG) in optical fibers, leading to spectrally flat and highly coherent super-continuum pulse [1, 2]. With its unique properties, parabolic pulses are also important in ultra-short pulse generation [2, 3], fiber amplifiers [4], optical regeneration[5], pulse retiming [6], spectral compression [7], and mitigation of linear waveform distortions [8], etc.. The triangular or sawtooth pulses are highly desired for many applications including time domain add-drop multiplexing [9], wavelength conversion [10], optical signal doubling [11], time-to-frequency mapping of multiplexed signals [12], etc.. In the field of particle accelerators, shaped laser pulse with flattop, in particular 3D ellipsoidal intensity distribution can significantly reduce the emittance of electron beams [13, 14], rendering much desired beam brightness improvement that is otherwise impossible. Besides those conventional pulse shapes, other laser pulse shapes are also useful, for example, double pulses with tunable temporal spacing and amplitude ratio can significantly improve laser ablation quality on metals[15].

Laser pulse shaping appears to be more difficult in the picosecond regime than other time regimes, as the widely used spectral-domain shaping such as grating-based shapers or acousto-optic programmable dispersive filter(DAZZLER) designed primarily for femtosecond pulses wouldn't work in this case due to the requirement on broader laser spectral bandwidth. On the other hand, pulse shaping from nanosecond to millisecond timescales is generally realized by acousto-optic or electro-optic systems which nevertheless are not fast enough for picosecond laser pulse shaping [16]. Several methods have been proposed for narrow spectrum picosecond pulse shaping, such as the use of superstructured fiber Bragg gratings [17], interferometer-like structures with beam splitters [18] and cascaded birefringent crystals of different length with two orthogonally polarized combs of pulses [13], just to name a few. However, none of these previously demonstrated techniques allow for practical optical arbitrary waveform generation (OAWG) for narrowband picosecond laser pulses. Will, et.al used multiple birefringent crystals and polarizers with Solc fan filter structure to generate flattop pulse [19]. Zhang, et.al recently reported a fiber-based arbitrary temporal picosecond laser shaping technique intended for laser material processing that once again involves strongly-chirped broadband seed pulses and relies on fairly complicated time-frequency manipulation. In addition, the liquid crystal controller can not be directly used for high power lasers [16]. There are also other efforts for temporal shaping of picosecond lasers [20, 21]. However, laser shaping by these methods is inflexible in the practice as it has to deal with precise adjustment of an array of different optical paths, and is sensitive to mechanical or thermal perturbations. Moreover, such kind of shaping methods are basically intended to produce only a few limited number of pulse shapes [22].

In this paper, we present a new pulse shaper that is based on Solc folded filter concept [23] to produce picosecond laser pulses with various profiles and linear polarization. The shaper consists of multiple birefringent crystals and polarizers for generating desired pulses with basically any predefined shapes through coherent pulse stacking. This shaping system features good long-term stability, is directly applicable to narrow spectrum picosecond pulses, easy to operate and to be automated, capable of shaping pulses over a wide wavelength range from IR to UV at any repetition rate and time structure, and is suitable for high power lasers without the need of manipulating the oscillators or amplifiers in the laser system, which tends to causes instability.

## 2. Experiment and results

Figure 1 shows the optical configuration of the variable laser pulse shaper. Two polarizers #1 and #2 are aligned with their polarization direction crossed to each other. Eight birefringent crystals of identical thickness are placed between polarizer #1 and #2. All of the crystals are a-cut, i.e. the optical axis of each crystal is parallel to its surface. The 8 crystals are arranged in such a way that the equation $\Theta_n = (-1)^n \frac{45^o}{N} + 90^o$ is satisfied, where $\Theta_n$ is the angle between the optical axis of the *n-th* crystal and the polarization direction of the input polarizer #1, $N$ is the total number of the crystals and $n(=1,2,\cdots,N)$ stands for the crystal sequence number. Each input pulse is split into two replica pulses after passing through one birefringent crystal, one being ordinary (o) light pulse and one extraordinary (e) light pulse, with a time delay introduced between o and e pulses due to their group velocity difference in the birefringent crystal. In total there are $2^8$ mutually delayed replica pulses after the 8-th crystal. After travelling through the output polarizer #2, these replica pulses interfere with each other and form a laser pulse with a pre-defined shape.

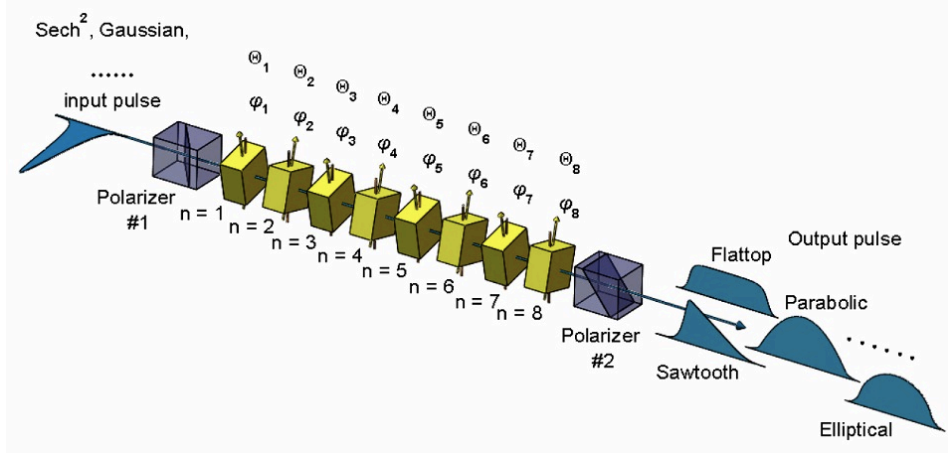

Fig. 1. The optical layout of the multiple birefringent crystal shaper used for pulse shaping in this paper. $\Theta_n$ is the angle between the optical axis of the *n-th* crystal and the polarization direction of the input polarizer #1. $\varphi_n$ is the crystal phase delay of the *n-th* crystal.

In this work, eight YVO$_4$ crystals were used for shaping 532nm picosecond laser pulses. The thickness of each YVO$_4$ crystal is 3.3mm. The end surfaces of the crystals were anti-reflection coated at 532nm with a residual reflection less than 0.1%. Each individual crystal is housed in a crystal oven and its temperature can be independently controlled from 40°C to 180°C with an accuracy of 0.1°C. Due to different thermal optical coefficients between o and e light beams and thermal expansion in the crystal, the crystal phase delay can be fine-tuned between $0 \sim 2\pi$ rad by adjusting the temperature of each crystal, which can strongly influence the interference among the replica pulses upon exiting the output polarizer #2. The phase delay of each single crystal, which determine the phases of the replica pulses, is defined as $\varphi = 2\pi(cT/\lambda - \lfloor cT/\lambda \rfloor)$, where symbol $\lfloor \ \rfloor$ means an operation that omits decimal and only keeps integer (e.g., $\lfloor 10.8 \rfloor = 10$), $T = l|1/v_{g,o} - 1/v_{g,e}|$ is the time delay between the o and e pulses after passing through one crystal, $c$ is the speed of light in vacuum, $\lambda$ is the laser wavelength, $l$ is the crystal length, and $v_{g,o}$ and $v_{g,e}$ are the group velocities of o and e pulses in a YVO$_4$ crystal, respectively. Therefore, the phase difference introduced between o and e beams after passing through one crystal is $\varphi + 2\pi \lfloor cT/\lambda \rfloor$. Each crystal oven is mounted on a kinematic rotation stage, providing pitch, yawn and continuous 360° rotation adjustments around the laser beam propagation axis. The input and the output polarizers are two identical Glan polarizers with an extinction ratio on the order of $10^5$:1.

Under the condition that all the crystals are of equal thickness and introduce the same time delay $T$, the $2^N$ replica pulses exiting the shaper can be arranged into $N + 1$ groups with equal group delays $m * T, m = 0, 1, \cdots, N$ [24], to form $N + 1$ replica pulses. Since YVO$_4$ is positive uniaxial crystal, the group velocity of the e beam is smaller than that of the o beam in the crystal. When crystal sequence number $n$ is an odd (even) number, increasing $\Theta_n$ would decrease (increase) the amplitude of the *n-th* and the *(n+1)-th* replica pulses at the same time increase (decrease) the amplitude of all other $N - 1$ replica pulses, and vice versa. Based on these rules, by rotating the crystals with respect to the laser propagation axis, the relative amplitudes of these $N + 1$ replica pulses can be flexibly adjusted to generate output pulses with basically any predefined shape.

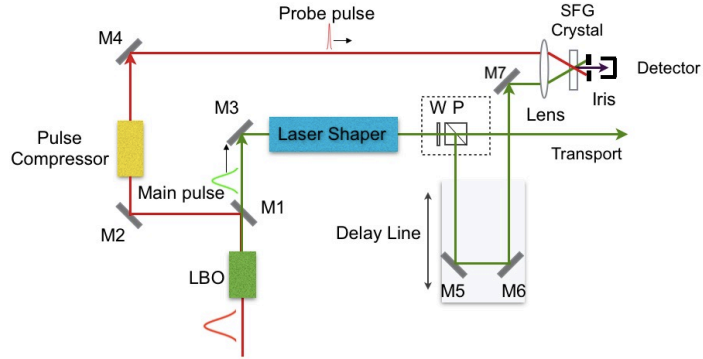

Fig. 2. Optical layout of the experimental setup including a cross correlator for pulse temporal profile measurement. M1: dichroic mirror，M2～M7：high reflectors, W: half-wave plate, P: polarizer (a variable beam splitter is formed by the combination of W and P).

Figure 2 depicts the experimental setup used for generation and measurement of the shaped pulses. The laser used in the experiment is a commercial fiber laser operating at 1064nm and a repetition rate of 8.125MHz. A LBO second harmonic generator converts the near IR beam to 532nm beam. The 532nm pulses possess a near Sech$^2$ temporal profile with 6.5ps (FWHM) pulse width before entering the pulse shaper. A portion of the residual 1064nm beam was coupled into a single mode polarization maintaining fiber to broaden the pulse spectral width (FWHM) from 0.26nm to about 5nm through the Self Phase Modulation (SPM) effect, and subsequently compressed by a grating compressor [25], resulting in much shorter sub-picosecond probe pulses needed for measuring the temporal intensity envelope of the shaped 532nm pulses by cross correlation between the shaped 532nm pulses and the precisely synchronized 1064nm probe pulses. The nonlinear crystal in the cross-correlator is a 2mm thick BBO crystal. The pulse width of the probe pulse was monitored by another commercial auto-correlator. The setup described here allows to measure the envelope of the shaped pulses in real time with sub-picosecond temporal resolution.

To achieve both high transmittance and high stability, the temperature of each crystal needs to be carefully adjusted to set each crystal phase delay close to $\pi$ before shaping the laser pulses, since the initial crystal phase delays of eight YVO$_4$ crystals are randomly distributed between $0 \sim 2\pi$ in the beginning. The output pulse from the shaper appears periodic with respect to the crystal phase delay. Within one period, increasing or decreasing all the crystal phase delays away from $\pi$ would reduce the beam transmittance through the shaper. Crystal phase delay near 0 rad would significantly reduce the shaper transmittance to nearly zero. In the experiment, a 3.3mm long YVO$_4$ crystal required a temperature variation of roughly about 36ºC to shift the crystal phase delay by $2\pi$ for 532nm laser. Since the crystal temperature control precision is 0.1ºC in our setup, the crystal phase delay can be adjusted with high resolution of about $0.0056\pi$.

The adjustment of rotation angles $\Theta_n$ ($n = 1, 2, \cdots\cdots, N$) for each crystal was accomplished in two steps. First, we rotated the crystal optical axes approximately to the calculated rotation angles as shown in Table 1 which gives the rotation angle corrections of the crystals for pulses of different shapes with all crystal phase delays set to $\pi$. Due to inevitable error in the crystal fabrication process such as the crystal cut angle, crystal length, inaccuracies on setting the crystal rotation angles and temperature, etc., the measured output pulse profile generally appeared close to the predefined shape, but not exactly the same. Nevertheless this provided a very good starting point and allowed us to further optimize each crystal rotation angle $\Theta_n$ to adjust the output pulse shapes until they match the pre-defined target shapes with the help of real time cross correlation measurement.

Table 1. Calculated rotation angle corrections $\Delta\Theta_n$ for different output pulse shapes, where $\Delta\Theta_n = \Theta_{n,tuned} - [(-1)^n \frac{45^o}{N} + 90^o]$. All the crystals in the shaper are a–cut YVO$_4$ with identical thickness of 3.3mm and phase delay set to $\pi$ rad. The input pulse profile is a Sech$^2$ distribution, with 6.5ps FWHM pulse width and 532nm wavelength.

| Crystal sequence number $\Delta\Theta_n$ (Deg) / Different shapes | 1 | 2 | 3 | 4 | 5 | 6 | 7 | 8 |
|---|---|---|---|---|---|---|---|---|
| Parabolic | 1.8 | -2.05 | 1.05 | -0.75 | 0.75 | -1.05 | 2.05 | -1.8 |
| Flattop | -1 | 0.4 | -4.05 | 5.45 | -5.45 | 4.05 | -0.4 | 1 |
| Elliptical | 1.65 | -2.1 | 2 | -1.52 | 1.52 | -2 | 2.1 | -1.65 |
| Triangular | 0.4 | -1.7 | -1.3 | 3.2 | -3.2 | 1.3 | 1.7 | -0.4 |
| Sawtooth-I | 1.4 | -4.4 | -1.2 | -2.7 | -2.5 | -1 | -1.3 | 2.2 |
| Sawtooth-II | -2.2 | 1.3 | 1 | 2.5 | 2.7 | 1.2 | 4.4 | -1.4 |

Several output pulse profiles measured after the shaper are shown in Figure 3, including parabolic, flattop, elliptical, triangular, and sawtooth shapes. The shaped pulse profiles are in good agreement with the predefined target profiles and are very smooth, which can be primarily attributed to the fact that the time delay $T = 3.4$ps of each crystal is much shorter than the input pulse width (6.5ps) and all the crystal phase delays were equally set to $\pi$. The total transmittance of the shaper typically is about 20%, depending on the specific output pulse shape. Taking into account of the reflection loss and absorption on each element in the system, the total transmittance could be over 30%, and our calculation shows it is even possible to reach 40~50% for most pulse shapes under optimal condition. We believe the crystal parameters including fabrication tolerance, etc., may be responsible for the loss. The intensity modulation in the flattop region of the flattop pulse is only about 0.92%(rms), which could be further reduced through a finer optimization of the crystal rotation angles. The leading and falling edges of the shaped pulses are primarily determined by the input pulse width and profile. One way to make the profile edge of the shaped pulse closer to the desired ideal geometrical shapes is to use shorter input pulse and more crystals although the pulse shapes in Figure 3 are satisfactory for many practical applications.

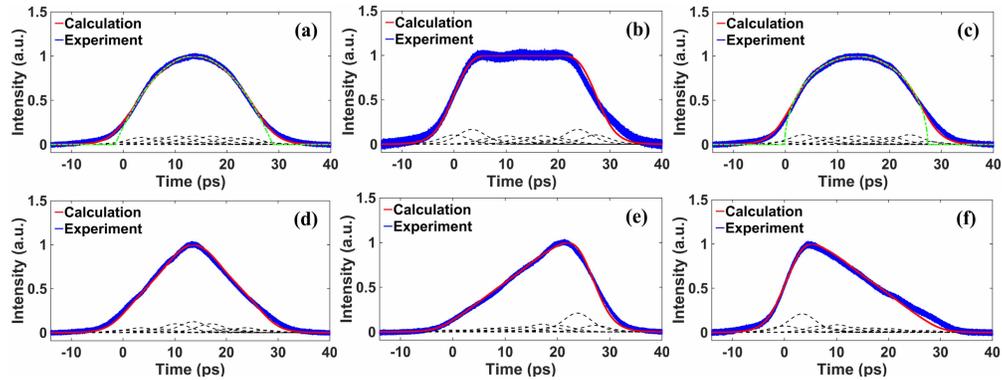

Fig. 3. Results of both the measured (blue lines) and the calculated (red solid lines, corresponding crystal rotation angle corrections are listed in Table 1) pulse profiles after the shaper, (a) Parabolic, (b) Flattop, (c) Elliptical, (d) Triangular, (e) Sawtooth-I, (f) Sawtooth-II. Black dashed curves are theoretically calculated 8+1 replica pulses. Green dash-dotted lines are the ideal parabolic and elliptical shapes. Birefringent elements in the shaper are identical. See detailed description in the text.

The stability is very important for pulse shaping system that determines its ultimate usefulness but was unfortunately seldom mentioned in the works previously reported by others in literature. By experience we know the pulse profiles from the popular birefringent crystal shaper scheme without temperature control tends to change noticeably over time [26] and has limited its applications in particular in scientific research. To check the shaped pulse stability of the shaper in our experiment, the phase delays of all the crystals were tuned to π by adjusting crystal temperature. The target pulse profile was then achieved by rotating the crystals with respect to the beam propagation axis. During a 8-hour continuous operation, the pulse shape stayed nearly constant as an example can be seen from the recorded data in Fig.4(a) for a triangular pulse. The beam transmittance through the shaper also remained unchanged, staying at about afore-mentioned 20% level, which indicates that the shaper has an excellent long-term stability. It was found this high degree of stability is primarily due to the fact that crystal phase delays were fine tuned and maintained through precise control of the crystal temperature, and were all set to π, which makes the system insensitive to the temperature fluctuation. Fig4(b) shows the calculated triangular pulse profiles with all the crystal phase delays set to $\pi, 0.98\pi, 0.96\pi, 0.94\pi, 0.92\pi$, and $0.9\pi$, which corresponds to crystal temperature changes by 0°C, 0.36°C, 0.72°C, 1.08°C, 1.44°C, and 1.8°C, respectively. The specific crystal rotation angle corrections for the calculated triangular pulse profiles are listed in Table 1. When the temperature of the crystal changed by 0.36°C, the relative peak intensity change with all the crystals phase delays set to $\pi, 0.98\pi, 0.96\pi, 0.94\pi$, and $0.92\pi$ are about 0.7%, 2.1%, 3.4%, 4.7%, and 6%, respectively. The effect of the crystal temperature fluctuation upon the shaped pulse is negligible when all the crystal phase delays are set to π. However, the shaped pulse profile changes more and more noticeably as crystal phase delay wanders away from π with the same amount of temperature variation, as shown in Fig.4(b). The reason behind the fact that the excellent temperature stability was achieved when all the crystal phase delays were set at π can be simply explained by the coherent stacking process as follows. The laser pulse intensity of the superposition of any two replicas $E_j(t)$ and $E_k(t)$ of the input pulse is $I_{jk}(t) = E_j^2(t) + E_k^2(t) + 2E_j(t)E_k(t)cos\,(\varphi'_j - \varphi'_k)$, where $E_j(t)$ and $E_k(t)$ are the fields of the j-th and k-th replicas of the input pulse ($j, k = 1,2,\cdots, 2^8$), and $\varphi'_j$ and $\varphi'_k$ are the phases of the j-th and k-th replicas, respectively. Assuming all the crystal phase delays ($\varphi$) are the same, the $\varphi'_j$, $\varphi'_k$ and phase difference $\Delta\varphi' = \varphi'_j - \varphi'_k$ would be the product of an integer and $\varphi$. When all the crystal phase delays are set to π, where 0 rad is not selected for which makes the beam transmittance of the shaper nearly to zero, the magnitude of the partial derivative of the $I_{jk}(t)$ with respect to $\Delta\varphi'$ would land at a minimum, i.e. zero. Therefore, at this point the influence of the phase difference ($\Delta\varphi'$) change induced by the temperature fluctuation of the crystal over the intensity of the stacked pulse is the smallest.

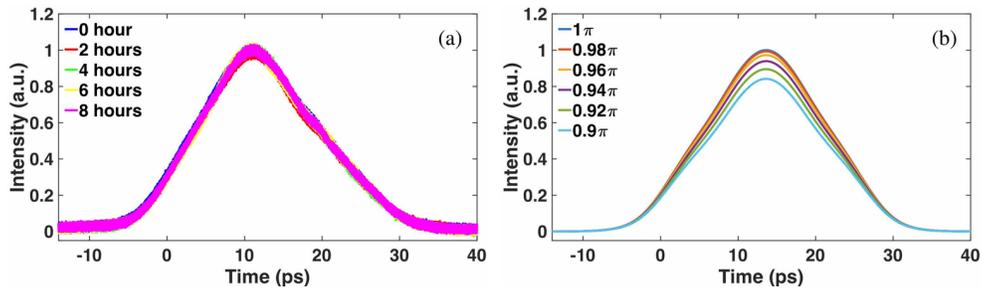

Fig. 4. (a) Recording of measured triangular pulse shapes at different times during 8-hour continuous operation when each crystal phase delay was set to π. (b) Calculated triangular pulse profiles with all the crystal phase delays set to $\pi, 0.98\pi, 0.96\pi, 0.94\pi, 0.92\pi$, and $0.9\pi$, respectively. The specific crystal rotation angle corrections for the calculated triangular pulse profiles are listed in Table 1. The crystals and the input pulse are the same as those in Table 1.

## 3. Further Discussion

With 8 crystals, many different shaped pulses can be produced stably and reproducibly with our shaper. To obtain more complicated pulse shapes, more crystals maybe needed, but would not noticeably increase the complexity of the adjustment process and operation of the system. It is necessary to mention that the YVO4 crystal is transparent from 0.4 to $5\mu m$. For UV laser shaping, other popular crystal such as BBO with extended shorter wavelength transparency down to $0.19\mu m$ may be chosen. According to the calculation, one way to further increase the beam transmittance through the shaper is to increase the ratio $T_1/T$, where $T_1$ is the pulse width of the input laser pulse.

For a given crystal phase delay variation $\Delta\varphi$, the corresponding crystal temperature variation is inversely proportional to the crystal length. Therefore, shapers using shorter crystal may also have better temperature stability. However, the actual crystal length needs to be considered together with other factors such as the pulse duration of the input pulse and output pulse, and the total number of crystals, etc..

Our simulation also reveals an interesting phenomenon as follows. If $\Theta_n$ ($n$ is the crystal sequence number) is denoted as the crystal rotation angle setting of the shaper system, then a shaper with $-\Theta_n$ crystal rotation angle setting would produce the same output pulse as those with $\Theta_n$ angel setting, as shown in Fig.5(a) and (c), or Fig.5(b) and (d), respectively. The shaper with $\Theta_n \pm 90°$ crystal rotation angle setting would generate mirrored output pulses with respect to those with $\pm\Theta_n$ angle setting, as shown in Fig.5(a) and (b), or Fig.5(c) and (d), respectively. This is equivalent to a situation when positive uniaxial crystals (e.g. YVO4) are replaced by negative uniaxial crystals (e.g. BBO). The specific crystal rotation angle values for the sawtooth pulses in Fig.5 are listed in Table 2. These characteristics show that the shaper can be configured into four different layouts to serve different application purposes.

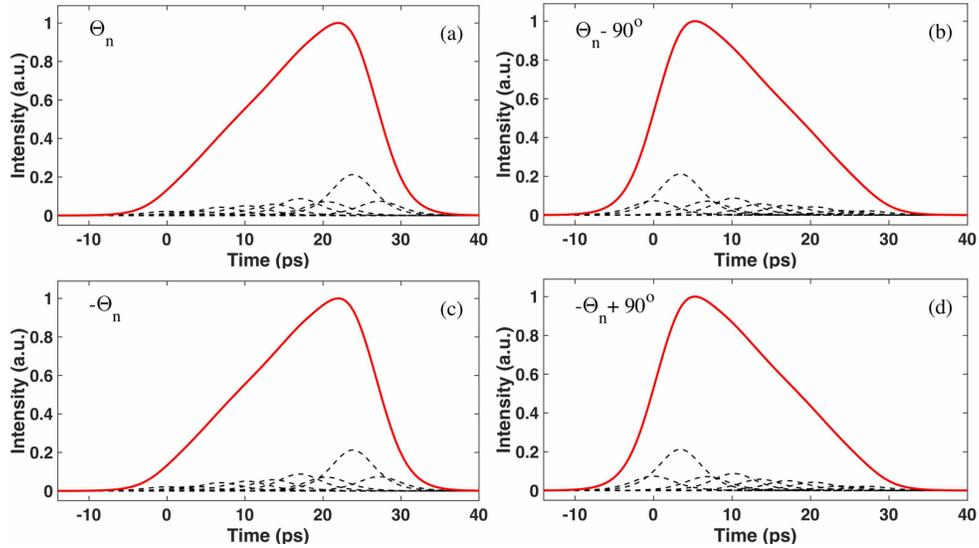

Fig. 5. Calculated (red solid lines) sawtooth pulses of the shaper with crystal rotation angles set at (a): $\Theta_n$, (b): $\Theta_n - 90°$, (c): $-\Theta_n$, and (d): $-\Theta_n + 90°$, respectively. The specific crystal rotation angle values for different sawtooth pulses in this figure are listed in Table 2. Black dashed curves are theoretically calculated 8+1 replica pulses. The crystals and the input pulse are the same as those in Table 1.

**Table 2. Calculated crystal rotation angles ($\Theta_{n,tuned}$) for sawtooth pulses of the shaper with crystal rotation angles set at $\Theta_n$, $\Theta_n - 90°$, $-\Theta_n$, and $-\Theta_n + 90°$, respectively. The crystals and the input pulse are the same as those in Table 1.**

| Crystal rotation angles $\Theta_{n,tuned}$(Deg) <br> Crystal sequence number | $\Theta_n$ | $\Theta_n - 90°$ | $-\Theta_n$ | $-\Theta_n + 90°$ |
|---|---|---|---|---|
| 1 | 85.775 | 85.775 - 90 | -85.775 | -85.775 + 90 |
| 2 | 91.225 | 91.225 - 90 | -91.225 | -91.225 + 90 |
| 3 | 83.175 | 83.175 - 90 | -83.175 | -83.175 + 90 |
| 4 | 92.925 | 92.925 - 90 | -92.925 | -92.925 + 90 |
| 5 | 81.875 | 81.875 - 90 | -81.875 | -81.875 + 90 |
| 6 | 94.625 | 94.625 - 90 | -94.625 | -94.625 + 90 |
| 7 | 83.075 | 83.075 - 90 | -83.075 | -83.075 + 90 |
| 8 | 97.825 | 97.825 - 90 | -97.825 | -97.825 + 90 |

From Table 1, we see that the above-mentioned single pass shaper can be converted into a new type of double pass shaper as shown in Fig.6, which has the advantage of drastically reducing the crystal quantity by half and ensuring the same symmetrical output pulse as those from the single pass shaper in Fig.1. If we follow the beam propagation direction and divide the total number of crystals into the first half (1 to 4) and the second half (5 to 8), for a double pass shaper the output polarizer and the second half of the crystals in the single pass shaper can be simply replaced by a retro-reflector and an quarter-wave-retarder (QWR). The fast or slow axis of the QWR should be aligned with the polarization direction of the input polarizer. The combination of retro-reflector and QWR work as a half-wave-retarder to rotate the polarization planes of all replica pulses and return the laser beam back on the same beam path. On the return pass, the replicas travelling backwards see the shaper the same way as they would go forwards through the second half of the crystals in the single pass shaper. The shaped pulse would then be conveniently kicked out by the same input polarizer upon its return. In other words, each crystal is used twice and only half number of crystals are needed, rendering both simplicity and cost reduction.

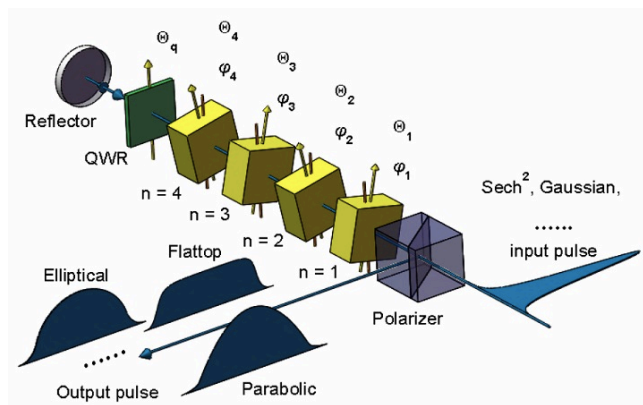

Fig. 6. The optical layout of an efficient double pass birefringent crystal shaper. $\Theta_n$ is the angle between the optical axis of the $n$-th crystal and the polarization direction of the input polarizer. $\varphi_n$ is the crystal phase delay of the $n$-th crystal. $\Theta_q$, representing the angle between the fast or slow axis of the QWR and the polarization direction of the input polarizer, is 0° for this double pass shaper.

The shaper described in this paper has distinctive features in contrast with other type of shaping devices. Due to the passive and static nature, it applies to lasers with any pulse repetition rate and time structure. The high damage threshold of the crystals and other optics used in the shaper allows laser pulse shaping under high power. Compared to typical pulse stacking methods [20, 26], the shaper reported here is insensitive to mechanical perturbations and temperature fluctuation, which is imperative for practical applications in particularly in high precision scientific research projects. This feature can be attributed to the fact that the fine tuning of the crystal phase delays is realized on an interferometric scale through precise control of the crystal temperature rather than splitting and combining multiple beams that are highly susceptible to misalignment [20, 21], and all the crystal phase delays are set to π. Although the initial input pulses have Sech$^2$ distribution, laser pulses with other shapes, such as Gaussian, Lorentzian, etc., can also be used with this shaper to generate laser pulses with various pre-defined temporal shapes. Moreover, the laser shaping scheme presented here is applicable to shaping even sub-picosecond laser pulses and to generating longer shaped laser pulses over 100ps. The system alignment and tuning of the crystals are relatively straightforward, the shaping process is highly reproducible and apparently can be implemented automatically by computer control with better precision and less time.

Another important feature of this shaper is that the shaped pulse from the shaper is linearly polarized with very high extinction ratio due to the use of Glan polarizers. Linear polarization beams are highly desired in many applications, such as laser amplification in amplifiers, devises that need polarizing elements, electron guns with photocathode drive lasers, and nonlinear optical processes including SHG, SFG, FHG, and so on.

## 4. Summary

In summary, a birefringent crystal shaper capable of generating variable shape picosecond pulses with linear polarization has been demonstrated and its essential characteristics were explored. Laser pulse with parabolic, flattop, elliptical, triangular, and sawtooth shapes were produced experimentally and agree well with the theoretical prediction. High transmittance and high stability can be achieved simultaneously by setting each crystal phase delay to π through fine control of the crystal temperature. Optical transmittance of ~20% was achieved and further improvement can be made according to the calculation. The shaper shows very good long term stability and is particularly suitable for narrowband picosecond laser pulse shaping. This shaping scheme is robust, easy to use and to be automated, capable of shaping laser pulses over a wide laser wavelength range from IR to UV, any pulse repetition rate, any time structure, and is directly applicable to high power lasers. This work may open up new opportunities for many potential applications. Finally, a practical double-pass temporal shaping configuration is proposed and could significantly reduce the structure complexity and the cost of this type of shaping systems.

**Funding**

National Key Research and Development Program of China (Grant No. 2016YFA0401904 and Grant No. 2017YFA0701000); National Natural Science Foundation of China (Grant No. 11735002).

**References**

1. S. Boscolo and C. Finot, "Nonlinear Pulse Shaping in Fibres for Pulse Generation and Optical Processing," International Journal of Optics **2012**, 14 (2012).
2. F. Parmigiani, C. Finot, K. Mukasa, M. Ibsen, M. A. F. Roelens, P. Petropoulos, and D. J. Richardson, "Ultra-flat SPM-broadened spectra in a highly nonlinear fiber using parabolic pulses formed in a fiber Bragg grating," Optics Express **14**, 7617-7622 (2006).

3. D. N. Papadopoulos, Y. Zaouter, M. Hanna, F. Druon, E. Mottay, E. Cormier, and P. Georges, "Generation of 63 fs 4.1 MW peak power pulses from a parabolic fiber amplifier operated beyond the gain bandwidth limit," Optics Letters **32**, 2520-2522 (2007).
4. M. E. Fermann, V. I. Kruglov, B. C. Thomsen, J. M. Dudley, and J. D. Harvey, "Self-Similar Propagation and Amplification of Parabolic Pulses in Optical Fibers," Physical Review Letters **84**, 6010-6013 (2000).
5. C. Finot, S. Pitois, and G. Millot, "Regenerative 40 Gbit/s wavelength converter based on similariton generation," Optics Letters **30**, 1776-1778 (2005).
6. F. Parmigiani, P. Petropoulos, M. Ibsen, and D. J. Richardson, "Pulse retiming based on XPM using parabolic pulses formed in a fiber Bragg grating," IEEE Photonics Technology Letters **18**, 829-831 (2006).
7. E. R. Andresen, J. M. Dudley, D. Oron, C. Finot, and H. Rigneault, "Transform-limited spectral compression by self-phase modulation of amplitude-shaped pulses with negative chirp," Optics Letters **36**, 707-709 (2011).
8. T. T. Ng, F. Parmigiani, M. Ibsen, Z. Zhang, P. Petropoulos, and D. J. Richardson, "Compensation of linear distortions by using XPM with parabolic pulses as a time lens," IEEE Photonics Technology Letters **20**, 1097-1099 (2008).
9. F. Parmigiani, P. Petropoulos, M. Ibsen, P. J. Almeida, T. T. Ng, and D. J. Richardson, "Time domain add–drop multiplexing scheme enhanced using a saw-tooth pulse shaper," Optics Express **17**, 8362-8369 (2009).
10. F. Parmigiani, M. Ibsen, P. Petropoulos, and D. J. Richardson, "Efficient All-Optical Wavelength-Conversion Scheme Based on a Saw-Tooth Pulse Shaper," IEEE Photonics Technology Letters **21**, 1837-1839 (2009).
11. A. I. Latkin, S. Boscolo, R. S. Bhamber, and S. K. Turitsyn, "Doubling of optical signals using triangular pulses," J. Opt. Soc. Am. B **26**, 1492-1496 (2009).
12. R. S. Bhamber, A. I. Latkin, S. Boscolo, and S. K. Turitsyn, "All-optical TDM to WDM signal conversion and partial regeneration using XPM with triangular pulses," in *2008 34th European Conference on Optical Communication*, 2008), 1-2.
13. A. K. Sharma, T. Tsang, and T. Rao, "Theoretical and experimental study of passive spatiotemporal shaping of picosecond laser pulses," Physical Review Special Topics - Accelerators and Beams **12**, 033501 (2009).
14. O. J. Luiten, S. B. van der Geer, M. J. de Loos, F. B. Kiewiet, and M. J. van der Wiel, "How to Realize Uniform Three-Dimensional Ellipsoidal Electron Bunches," Physical Review Letters **93**, 094802 (2004).
15. R. Le Harzic, D. Breitling, S. Sommer, C. Föhl, K. König, F. Dausinger, and E. Audouard, "Processing of metals by double pulses with short laser pulses," Applied Physics A **81**, 1121-1125 (2005).
16. B. M. Zhang, Y. Feng, D. Lin, J. H. V. Price, J. Nilsson, S. Alam, P. P. Shum, D. N. Payne, and D. J. Richardson, "Demonstration of arbitrary temporal shaping of picosecond pulses in a radially polarized Yb-fiber MOPA with > 10 W average power," Optics Express **25**, 15402-15413 (2017).
17. P. Petropoulos, M. Ibsen, A. D. Ellis, and D. J. Richardson, "Rectangular Pulse Generation Based on Pulse Reshaping Using a Superstructured Fiber Bragg Grating," J. Lightwave Technol. **19**, 746 (2001).
18. H. Tomizawa, H. Dewa, H. Hanaki, and F. Matsui, "Development of a yearlong maintenance-free terawatt Ti: Sapphire laser system with a 3D UV-pulse shaping system for THG," Quantum Electronics **37**, 697-705 (2007).
19. I. Will and G. Klemz, "Generation of flat-top picosecond pulses by coherent pulse stacking in a multicrystal birefringent filter," Optics Express **16**, 14922-14937 (2008).
20. Y. Park, M. H. Asghari, T.-J. Ahn, and J. Azaña, "Transform-limited picosecond pulse shaping based on temporal coherence synthesization," Optics Express **15**, 9584-9599 (2007).
21. J. A. Fülöp, Z. Major, B. Horváth, F. Tavella, A. Baltuška, and F. Krausz, "Shaping of picosecond pulses for pumping optical parametric amplification," Applied Physics B **87**, 79-84 (2007).
22. R. A. Meijer, A. S. Stodolna, K. S. E. Eikema, and S. Witte, "High-energy Nd:YAG laser system with arbitrary sub-nanosecond pulse shaping capability," Optics Letters **42**, 2758-2761 (2017).
23. I. Šolc, "Birefringent Chain Filters," J. Opt. Soc. Am. **55**, 621-625 (1965).
24. S. E. Harris, E. O. Ammann, and I. C. Chang, "Optical Network Synthesis Using Birefringent Crystals.* I. Synthesis of Lossless Networks of Equal-Length Crystals," J. Opt. Soc. Am. **54**, 1267-1279 (1964).
25. A. M. Johnson, R. H. Stolen, and W. M. Simpson, "80× single‐stage compression of frequency doubled Nd:yttrium aluminum garnet laser pulses," Applied Physics Letters **44**, 729-731 (1984).
26. S. Zhang, S. Benson, J. Gubeli, G. Neil, and G. Wilson, "Investigation and evaluation on pulse stackers for temporal shaping of laser pulses," in *Proc. 32 nd Int. Free Electron Linac Conf*, 2010, 394-397.